\documentclass[sigconf]{acmart}

\usepackage{tabularx}
\usepackage{multirow}
\usepackage{graphicx}
\usepackage{booktabs}
\usepackage{caption}
\usepackage{subcaption}
\usepackage{array}
\usepackage{float}
\usepackage{enumitem}
\usepackage{amsmath}
\usepackage{etoolbox}
\usepackage{xparse}
\setcopyright{none}
\renewcommand\footnotetextcopyrightpermission[1]{} % removes footnote with conference information in first column

\AtBeginEnvironment{tabular}{\small}

\newcommand{\paragraphb}[1]{\vspace{0.03in}\noindent{\bf #1} }
\newcommand{\subscript}[2]{$#1 _ #2$}
\newcolumntype{S}{>{\centering\arraybackslash}m{.5in}}

\def\ExplainVariable#1:#2:#3!!!{$#1$ &$\;\;=$ #2 \\}

\NewDocumentEnvironment{formula}{ t! O{where:} m b }{%
 \begin{equation}#3\end{equation}
  #2 \IfBooleanT{#1}{\\[1ex]}
  \renewcommand*\do[1]{\ExplainVariable##1::!!!}%
  \tabular[t]{r@{}l}\docsvlist{#4}\endtabular
}{}
\NewDocumentEnvironment{formula*}{ t! O{where:} m b }{%
 \begin{equation*}#3\end{equation*}
  #2 \IfBooleanT{#1}{\\[1ex]}
  \renewcommand*\do[1]{\ExplainVariable##1::!!!}%
  \tabular[t]{r@{}l}\docsvlist{#4}\endtabular
}{}

\begin{document}

%\title{Can You Still See Me?:\\Identifying TLS Encrypted Robot Operations using Traffic Analysis}
\title{Can You Still See Me?: Reconstructing Robot Operations Over End-to-End Encrypted Channels}
% Ryan: shall we say 'identifying robot operations over end-to-end encrypted channels' in the title?

\author{Ryan Shah, Chuadhry Mujeeb Ahmed, Shishir Nagaraja}
%\institute{University of Strathclyde, Glasgow}
\affiliation{%
	\institution{University of Strathclyde, Glasgow}
	% \city{Glasgow}
	\country{UK}
}
\email{{first.last}@strath.ac.uk}

\begin{abstract}
Connected robots play a key role in Industry 4.0, providing automation and
higher efficiency for many industrial workflows. Unfortunately, these robots
can leak sensitive information regarding these operational workflows to
remote adversaries. While there exists mandates for the use of end-to-end
encryption for data transmission in such settings, it is entirely possible for
passive adversaries to fingerprint and reconstruct entire workflows being
carried out -- establishing an understanding of how facilities operate. In this
paper, we investigate whether a remote attacker can accurately fingerprint robot
movements and ultimately reconstruct operational workflows. Using a neural
network approach to traffic analysis, we find that one can predict TLS-encrypted
movements with around \textasciitilde60\% accuracy, increasing to near-perfect
accuracy under realistic network conditions. Further, we also find that attackers
can reconstruct warehousing workflows with similar success. Ultimately, simply
adopting best cybersecurity practices is clearly not enough to stop even
weak (passive) adversaries.

% \comment{The importance of privacy is paramount. For example, identifying what operations a surgical robot is
% carrying out on a patient could reveal what procedure is being undertaken. If
% this is coupled with other patient meta-information, it could lead to potential
% compromise of the patient's privacy. Several studies have explored fingerprinting
% traffic, such as in the context of websites and users over TLS, yet little
% attention has been paid to the capabilities of a passive attacker in a robotics
% setting. In this work, we investigate whether a capable passive adversary can
% fingerprint the movements of a teleoperated robot, through the monitoring of
% TLS-encrypted traffic flows between the robot and its controller. We find that
% simply observing the flow patterns at a high level does not provide enough
% information to the adversary. Instead, taking a deep learning approach, }

\end{abstract}



\settopmatter{printfolios=true}
\maketitle
\pagestyle{plain}

% Introduction
\section{Introduction}
\label{sec:intro}

The field of robotics has seen a rise in implementations for a wide variety of
sectors, including the automotive and healthcare industry, bringing promise of
higher levels of accuracy and minimised risk of liability and arising
complications~\cite{schulz2007results,reutersrobots}. Whilst robotic systems
traditionally pose many safety challenges, being Internet-connected with networks
of sensors and other embedded devices exposes them to attacks in the cyber domain,
such as fingerprinting attacks~\cite{bezawada2018behavioral,msadek2019iot}.

Teleoperated robotics systems are becoming more prominent in a wide array
of environments, most notably in industry (i.e. warehouse robots~\cite{benali2018dual,rey2019human})
and healthcare (i.e. surgical robots~\cite{hannaford2012raven,tewari2002technique}).
In most teleoperated robots we observe a similar architecture. This consists of
a controller (or pendant) that is connected to the robot and provides feedback
to the (human) operator, as well as allowing complete control or selection of
pre-defined tasks.

While prior art has demonstrated several types of attacks on teleoperated robotic systems,
ranging from active targeted attacks such as modifying/dropping commands
in-flight and modifying feedback to operators~\cite{bonaci2015make,alemzadeh2016targeted,quarta2017experimental,demarinis2018scanning}, there has been less focus on
reconnaissance aspects. Although eavesdropping and fingerprinting is a more passive
opportunity for attackers, the resulting attacks on robotic operations
are still of importance. Even if strict security requirements are conformed to,
such as using cryptographic measures to ensure the confidentiality
of transmitted data, there are still arising threats that cannot be ignored. The
question we aim to address is whether it is still possible for an adversary
to infer information about what operations a robot is carrying out? Further, can
this be successful even when the communication is protected by appropriate
channel security technology (e.g. TLS)?
% Furthermore, alongside some key uses of robotic systems there may be privacy
% concerns that arise~\cite{butler2015privacy,lutz2020robot}. For example, let us assume that a typical surgical
% operation consists of a set of defined movements (i.e. in sinus surgery~\cite{baban2020surgical}). If an adversary is able to
% identify the robots movements, then detecting a pattern of a subset of these
% movements could leak information about the surgical procedure being carried out.
% From here, the adversary could gather other information such as patient admission
% and exit times, among other things, which could lead to a breach of patient
% privacy.

\paragraphb{End-to-End Encryption Measures.} While it is common sense to ensure
the integrity/confidentiality of data transmission in Internet-connected robotics
systems, in industrial contexts the use of end-to-end encryption is not mandated
nor is it defined as a standard protection measure. Interestingly, in other areas
such as surgical robots, the use of TLS is required for legal compliance as set
out by legislation such as HIPAA~\cite{hipaatls,annas2003hipaa,mercuri2004hipaa}
and MDR~\cite{mdr2017}. In any case, however, although TLS is used to protect
the confidentiality of transmitted data, we find in this work that this is not
enough to ensure confidentiality of operational workflows.

% The use of secure channels like TLS for surgical robots is required for
% legal compliance. The Health Insurance Portability and Accountability Act
% (HIPAA) states that electronic protected health information (PHI) should be
% encrypted in accordance with the HIPAA security rule which requires that ``the
% use of an algorithmic process to transform data into a form in which there is a
% low probability of assigning meaning'' is a requirement and should ``comply, as
% appropriate, with NIST 800-52, 800-77 and 800-113 [...] among other Federal
% Information Processing Standards (FIPS) 140-2'' validated specifications~\cite{hipaatls}.
% Although, TLS is used to prevent unauthorised access or disclosure of information,
% attackers may apply our operation reconstruction attack on traffic to break patient
% privacy. This could lead to {\em potential} HIPAA violations. This includes (but is not limited to):
% unauthorised access to and release of PHI to those whom of which are not
% authorised; impermissible disclosures of PHI; and theft of patient
% records~\cite{annas2003hipaa,mercuri2004hipaa}.

\paragraphb{Our Contributions.}
In this work, we explore the threats to the privacy of the
physical processes and people involved in the outcomes of such an attack. The
key idea is to extract unique traffic features on the movements of a teleoperated
robot from the network layer data. For our experiments, we collected samples of
TLS-encrypted traffic by eavesdropping on the communication channel between a
robot and controller in a teleoperated architecture. By analysing this traffic, we
show that an adversary could pin-point individual movements, however combinations
of movements are not so easy to infer. This provoked a response to use a
neural network, which has shown to be successful in other areas such as website/user
fingerprinting~\cite{oh2019p1,abe2016fingerprinting}. By using a neural network, we show that
an adversary can identify movement patterns averaging around \textasciitilde 60\%
accuracy, reaching perfect accuracy under realistic network conditions of variable packet loss and link delay. Furthermore,
we demonstrate that this adversary can also reconstruct warehousing workflows with
high accuracy. To conclude, we also investigate the use of Tor as a countermeasure
against our proposed attack.

\paragraphb{Organisation.}
The remainder of the paper is organised as follows: In Section~\ref{sec:background},
we provide a more in-depth background on robotic systems and
the problem space. In Section~\ref{sec:analysis}, we discuss the environment
in which we conducted traffic analysis and detail our findings. In
Section~\ref{sec:discussion} we provide a discussion of this work, and related
work is described in Section~\ref{sec:related}. We then conclude in
Section~\ref{sec:conclusion}.

% Background
\section{Background}
\label{sec:background}

\subsection{Teleoperated Robotics}

\begin{figure}[h]
	\centering
	\includegraphics[width=0.75\linewidth]{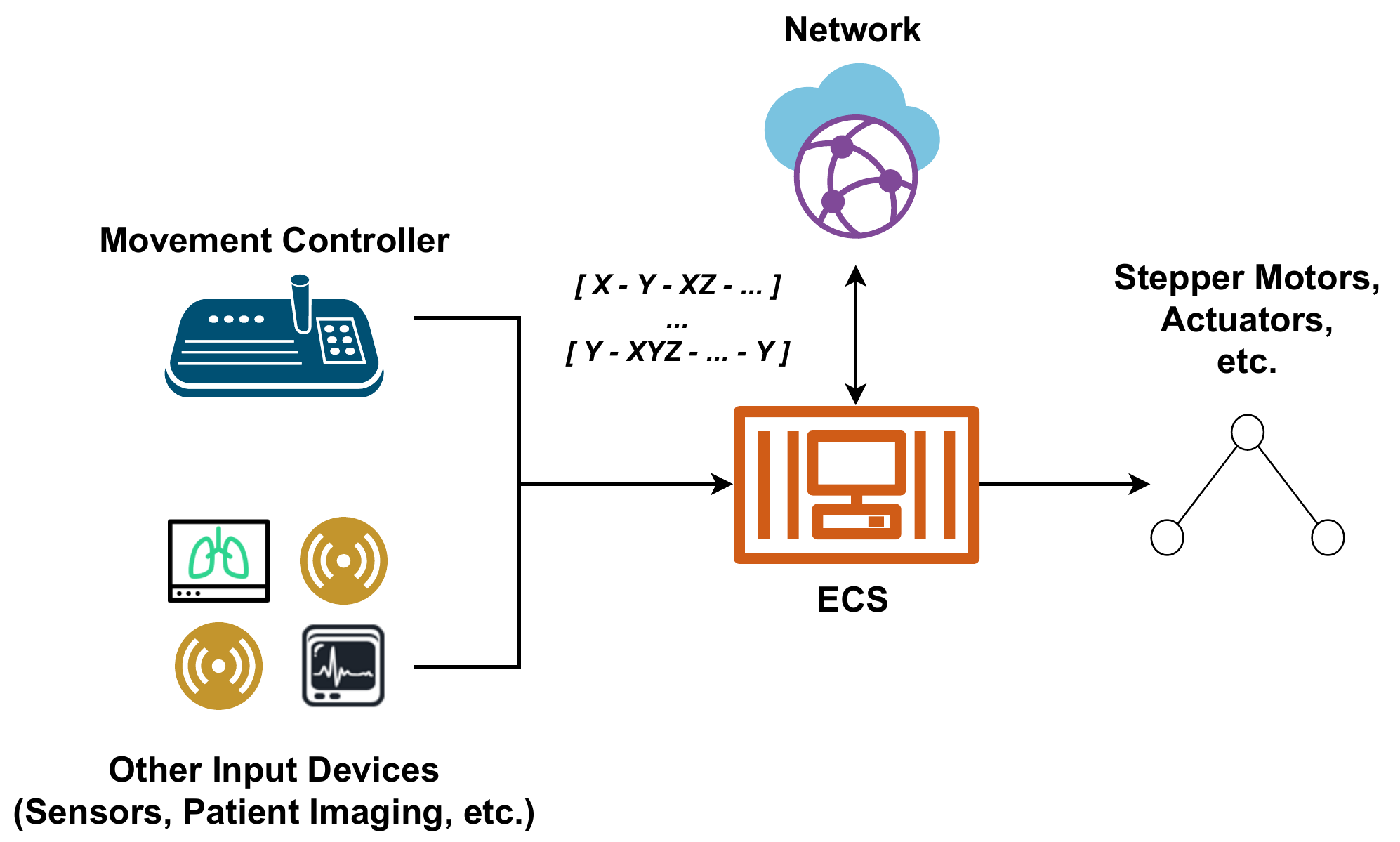}
	\caption{Typical Teleoperated Robot Architecture}
	\label{fig:robotarch}
\end{figure}

The use of robotic systems in safety-critical environments has increased,
with promise of higher levels of accuracy and precision of operation, and ultimately
minimising the risk of complications that arise during and after operation.
Reuters describe that the shipments for robotic systems has increased by
nearly 16\% from 2017, which ultimately corresponds to an increase in
robotic installations~\cite{reutersrobots}. While previous robotic suites
were configured based on pre-planned operations~\cite{bargar1998primary,schulz2007results},
many modern robotic suites are
teleoperated~\cite{dalvand2014improvements,alemzadeh2016targeted,bonaci2015make,tewari2002technique}.
These systems share a common architecture~\cite{quarta2017experimental}, consisting
of a controller (e.g. a surgeon's console or teach pendant), a set of input and
output devices (e.g. sensors and actuators) and a network in which the robot
operates, linked together via an electronic control system (Figure~\ref{fig:robotarch}).

The link between the controller and main control system is a
key safety-critical component of system operation, where feedback
from the robot and inputs to the robot traverse. The type of controller
differs between operational contexts, however the fundamental use of the
controller (i.e. to send control commands) remains the same.
For example, in surgical robotic systems the controller has evolved into a
console~\cite{tewari2002technique,hannaford2012raven} which is operated by
a human who uses finger controllers and foot pedals that translate
human movements into instructions the robot can interpret. In contrast,
industrial robots are equipped with a teach
pendant~\cite{quarta2017experimental,dalvand2014improvements}
that consists of touch screen with an interactive GUI or physical
buttons to control the robot.

\subsection{Threat Model}
%\paragraphb{Threat Landscape}

\begin{figure}
  \centering
  \includegraphics[width=0.825\linewidth]{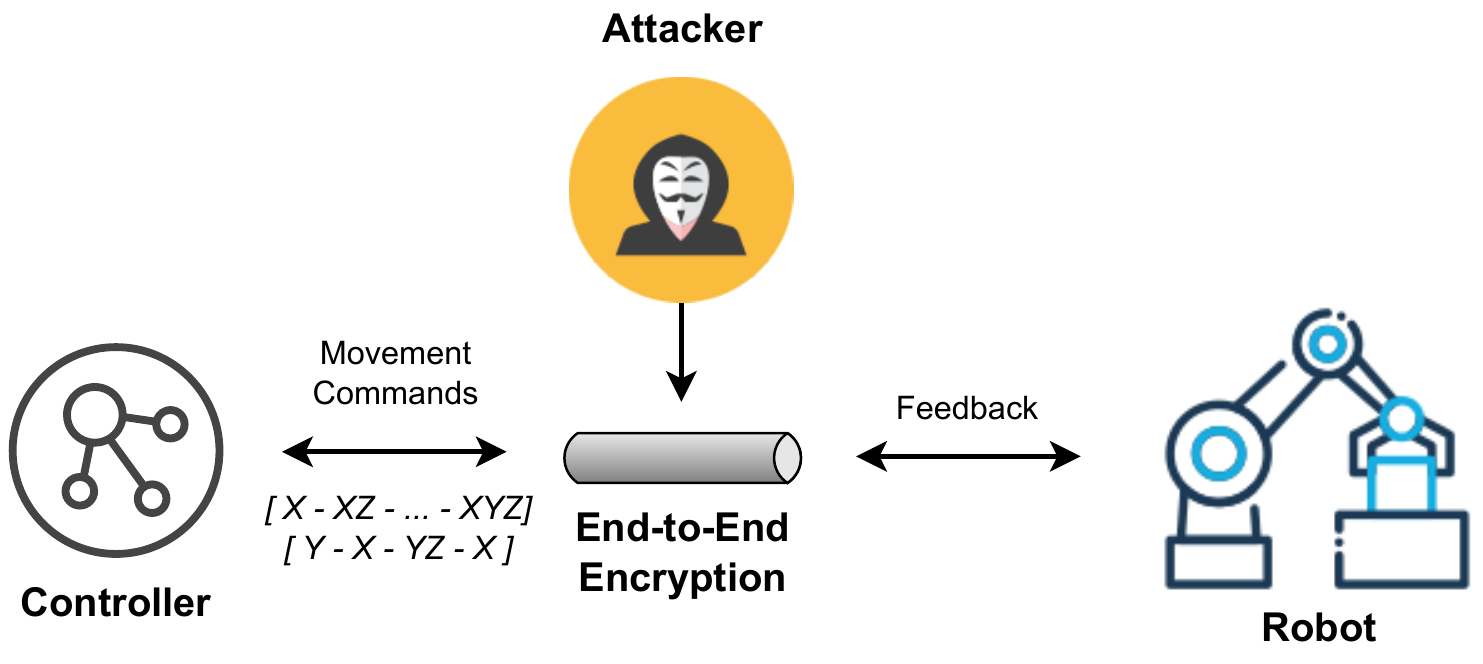}
  \caption{Robot and Traffic Analysis Setup}
  \label{fig:setup}
\end{figure}

Although several defenses and safety mechanisms have been
proposed~\cite{bonaci2015make,alemzadeh2016targeted,quarta2017experimental,demarinis2018scanning},
many attacks focus on an active adversary, with little attention paid to weaker
(capabilities of) passive adversaries and their impact on operational privacy.
Here, we consider a passive adversary to be able to eavesdrop on the
communication channel between the key component of modern robotic systems,
the controller, and the robot itself (Figure~\ref{fig:setup}).
First, an insider (e.g. technical staff who can access the internal organisational
or robot network) can potentially eavesdrop on robot operations. By collecting
the observed traffic and combining this with other data, information leakages
about the operational environment exposes another degree of detail. The technical
staff could potentially be bribed to sell this data on, or even possibly
attempt to discredit the operator or company, at the expense of
breaching operational privacy. Second, in the case of a passive outsider,
competing organisations could use operational information to compare
performance or, further, also use this information to compromise operational
privacy. In either case, we investigate whether it is possible for an adversary
to identify robot movements even when appropriate channel security is employed.
% Notably, this is a key requirement to prevent violation of the Health Insurance
% Portability and Accountability Act (HIPAA), where protected health information of
% any kind (i.e anything related to patients which {\em should} also include robotic
% systems) must be encrypted (via TLS) to prevent unauthorised access or disclosure.
% By identifying movements, an adversary could reconstruct known surgical
% procedures, and in combination with other sources of protected health information
% (such as that of patients operated on by a surgical robot) this could compromise
% PHI.

\paragraphb{Hypotheses and Goals}
Prior art has demonstrated traffic analysis over plain-text robot
traffic~\cite{mcclean2013preliminary}, or compromising privacy by tracking
system users~\cite{rodriguez2018message}, with security solutions such as
ciphering messages and encrypting network communication using
SSL/TLS~\cite{white2016sros}. However, the question remains as to whether it is
still possible to identify a robot's movements even when measures such as TLS
are employed? In this work, our primary goal is to determine whether an
adversary can infer robot movements from encrypted traffic flows alone. Second
to this, we hypothesise whether a higher level of granularity is present in this
case with potential reconstruction of procedures. By doing this, it is entirely
possible to pinpoint operational patterns and ultimately piece together entire
workflows.

\section{System Design}
\label{sec:analysis}

To these hypotheses and goals, we investigate whether an adversary can identify
our robot's operations and ultimately reconstruct entire workflows. The focus of
our study is on modern teloperated robot architectures, where the key
communication channel is between the controller (i.e. a teach pendant) and the
robot itself, as shown in Figure~\ref{fig:setup}. We made use of uFactory's
uARM Swift Pro, which is operated by an Arduino Mega 2560 % ~\cite{uarmswiftpro}
running MicroPython. The robot is connected to a controller to mimic the
teleoperated robot architecture, which is run on a Windows 10 laptop using the
uARM Python (3.8.X) SDK. In accordance with communications used for teleoperated
robots~\cite{zeng2018network}, we follow a master-slave communication topology
in which an asynchronous TCP/IP socket is established for peer-to-peer
communication. The message structure in our implementation involves only a
change to the robot system message (payload) in which system status and
position information is to be transmitted. The session management data class
in prior work is handled via TLS session establishment. To enable TLS-encrypted
communications between the robot and controller, we route traffic through a
software-defined network (SDN) using Mininet 2.3.0, running a TLSv1.2
client-server network for simulating realistic network conditions. Ultimately,
this protocol can be generalised to teleoperated robots in general due to
similarities they share in their architecture. The key generalisation here in
terms of the packet structure is movement data. Additional information such as
imaging, system information and sensor data do not directly contribute to
movement inference in non-autonomous systems, beside influence on human
operators to make decisions on these movements. Further, while our robot only
has 3 degrees of freedom (for position), additional degrees of freedom will only
increase payload. This may slightly influence inter-arrival and round-trip times.
In any case, this would be interesting to see the impact as a point of future
work for this attack. Finally, we used Wireshark to capture the encrypted
communication between the robot and controller, imitating our passive adversary.

There are several challenges to carrying out procedure reconstruction:
\begin{enumerate}[label=(\subscript{RQ}{{\arabic*}})]
	\item Can we detect individual robot movements on each axis?
	\item Can we detect combinations of robot movements?
	\item Is it possible to reconstruct operational workflows that correspond to
	      a set pattern of movements?
	\item How is the identification of movements affected by the distance and
	      speed in which the robot moves?
	\item How do realistic network conditions such as network delay affect
	      the identification of robot movements?
\end{enumerate}

\subsection{Challenges in Applying Traffic Analysis Approaches}
In order to understand the challenges better, we carry out a preliminary
analysis of the problem setting. In this work, the focus is on teleoperated
industrial robots and the reproduction of warehousing operations via the traffic
analysis side channel. We believe that our robot and network setup is a suitable
replication candidate for a traffic analysis study, in comparison with the likes
of other single-arm robots (of at least 3 degrees-of-freedom) used in
warehouses. This is because the majority of these robots consist of the same
principal components as our robot for movement inference (i.e. stepper motors,
actuators, etc.). Ultimately, the only meaningful data for movement
fingerprinting from a traffic analysis perspective is the transmission of
control commands which would affect timing patterns and message payload, among
others. Data from other components, such as sensing equipment, would be
indistinguishable to data provided by actuators and motors which operate via
control commands sent by the controller. However, these other components may
provide useful information from other side channels, such as acoustic or power,
which can be used in combination with the traffic analysis side channel.
Ultimately, the first step to addressing our challenges is to generate
appropriate traffic traces which will clarify the challenges involved in
applying traffic analysis approaches.

\paragraphb{Robot Traffic Traces.}
\label{sec:dataset}
We follow an emulation approach to generate a variety of traffic behaviour.
We programmed the uARM robot to carry out a set of movement operations it allows
along the X, Y and Z axes. In addition to capturing single movements, we also
programmed combined movements (i.e. movement along the X and Y axes
simultaneously). As well as combined movements, for more fine-grained analysis
we also programmed movements with varying distances (in mm), as well as varying
speeds of movement (mm/s) for a more fine-grained analysis. Finally, we wanted
to determine whether an adversary can fingerprint these movements under
realistic network conditions. The justification for our choice of network
emulation framework is given in Section~\ref{sec:discussion}. Within the
emulated network, we programmed varying network link delays (between the
controller and robot) and packet losses over this link. The dataset for the
range of experimental parameters contains around 150,000 samples.

Within our dataset, we collected the traffic features for packets travelling
between the robot and controller in the teleoperated architecture. Within each
packet sent from the controller to the robot, the payload data contains the set
of G-code instructions for the corresponding robot operation(s) to be performed.
From our traffic capture, we look at the following features: {\em Packet Time},
{\em Frame Length}, {\em FrameCapLen} (frame length stored in the capture file),
{\em IP Length}, {\em TCP Length}, {\em Bytes In Flight}, {\em Push Bytes Sent}
(bytes sent since last PSH flag), {\em ACK Round-Trip-Time (RTT)}, and
{\em TLS Record Length}.

\begin{figure}
  \centering
  \includegraphics[width=1\linewidth]{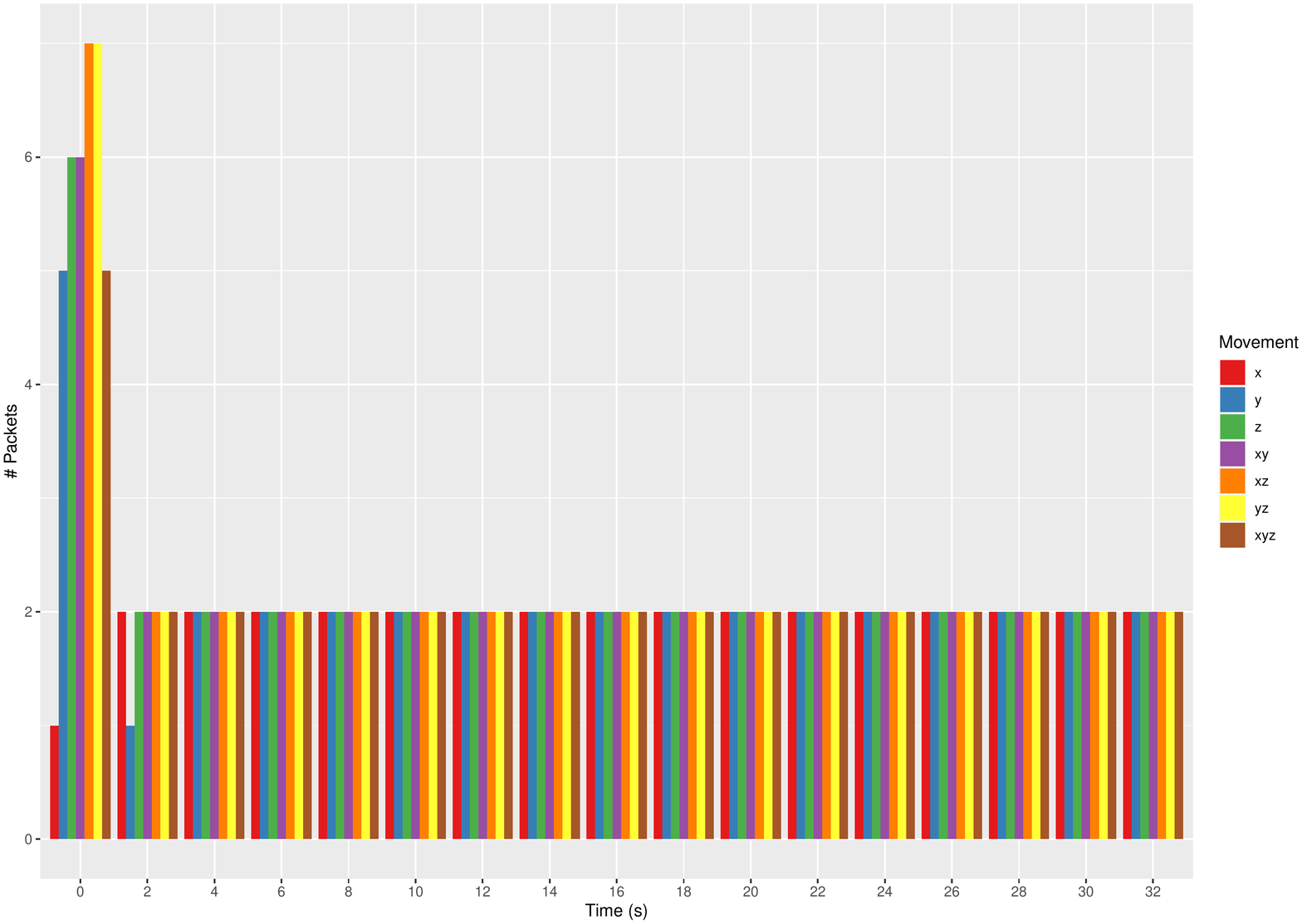}
  \captionsetup{singlelinecheck=off}
  \caption{\centering Packets per Second for Robot Movements\hspace{\textwidth}{\small\em\textmd{The flow
  of packets per second across movements show that time-series analysis alone does not provide enough basis
  for fingerprinting movements}}}
  \label{fig:timepacketspersec}
\end{figure}

\paragraphb{Packet Feature Analysis.}
First, we wanted to determine how {\em time-frequency} representations of
traffic features may offer clues towards what robot movement is being carried
out (Figure~\ref{fig:timepacketspersec}). We note that among all movements, an
average of 2 packets are sent between the robot and controller per second. The
second interval between packets sent corresponds to the programmed movement
interval in order to capture each movement's traffic data properly. We observe
an increase in packets at the start of each operation, which corresponds to the
initialisation of the communication between the robot and controller (such as
TLS initialisation and robot-controller configuration) and not the movements,
and thus are discounted. We found that the frequency of packets across time does
not vary past the initialisation at the beginning of each data collection step.

\begin{figure}
  \centering
  \includegraphics[width=1\linewidth]{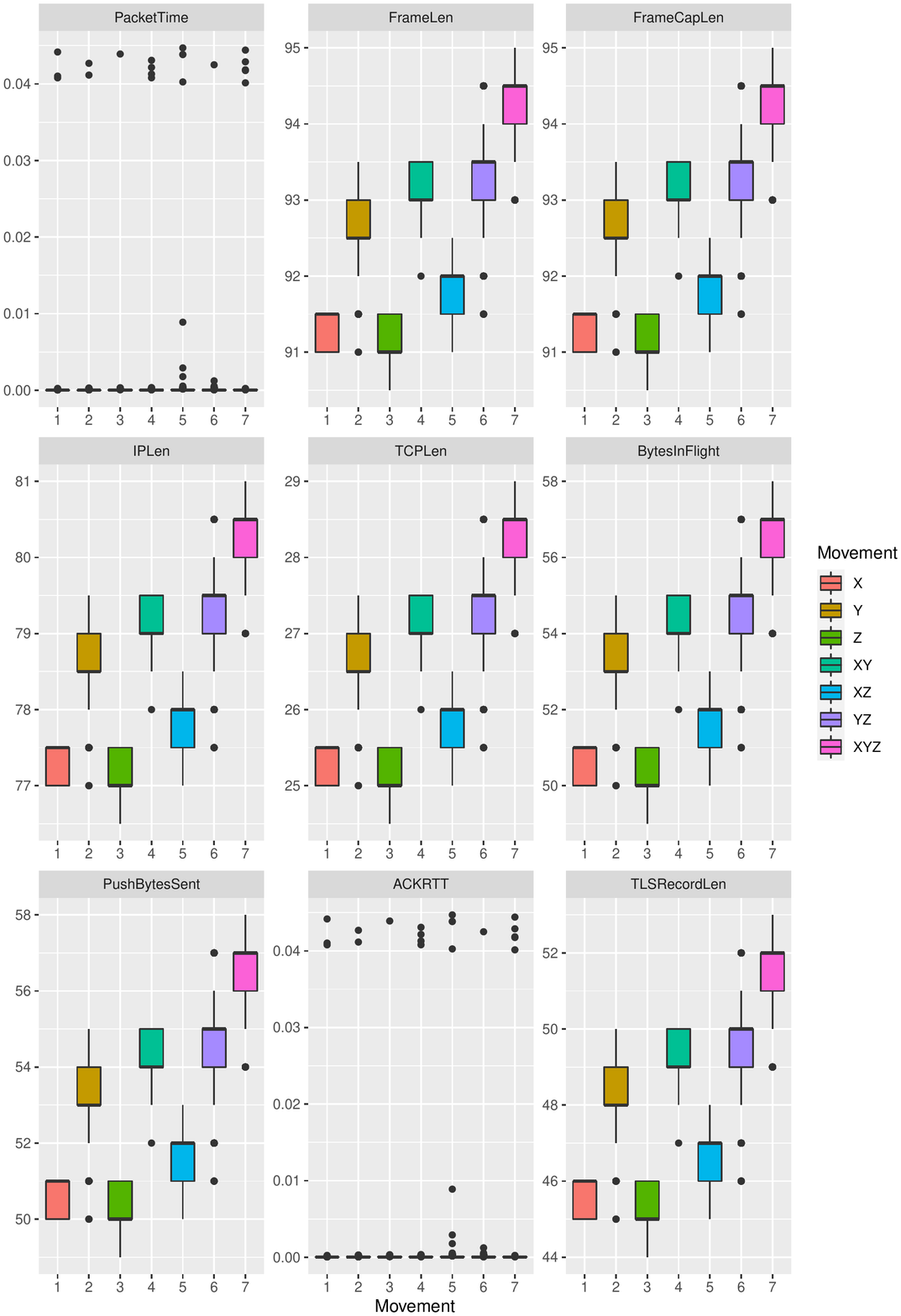}
  \captionsetup{singlelinecheck=off}
  \caption{
    \centering
    Traffic Features Across Movements
%     \hspace{\textwidth}{\small\em\textmd{A closer look shows
% 	there are more subtle variations across movements aside from only time
% 	variations}}
  }
  \label{fig:tfeatures}
\end{figure}

We then explored how each of the traffic features contained within these packets
vary across our movement samples (Figure~\ref{fig:tfeatures}). Here, the x-axis
is the movement and the y-axis is the size of the traffic feature. Initially,
across most of the features (aside from Packet Time and RTT), we observe a
similar pattern, with the XYZ movement being the most {\em identifiable} in
comparison. Both the X and Z movements appear to be more similar, also seen with
the XY and YZ movements, however the median values in each case are at opposite
ends of the inter-quartile range (which is the same for both sets of movements
in this comparison, respectively). Interestingly, the packet time and RTT are
relatively similar with little difference in the case of outlying values. For
the uARM robot, each packet and its corresponding traffic features constitute a
flow (each movement operation corresponds to a flow). In other robotic systems,
this flow may be split up and is a consideration for future work. Overall, we
found that simple approaches such as {\em eye-balling the dataset} or {\em basic
frequency analysis} is not enough to answer the challenges listed above.

\subsection{Workflow Reconstruction Approach}
\label{sec:recon}
For our reconstruction approach we chose to take a neural network approach to
recognise traffic features, identify robot movements and thereby reconstructing
warehousing workflows. This is based on a set of features from the encrypted
robot traffic. The effectiveness of applying machine learning techniques on
encrypted traffic has been demonstrated in other applications such as
VoIP~\cite{lotfollahi2020deep,wang2017end,voiploc}. Our reconstruction technique
involves several stages. First, we pre-process the data to remove noise. Second,
we identify basic traffic features. Next, we construct a neural network to
classify the features into robot behaviours in order to fingerprint encrypted
traffic flows thus aiding workflow reconstruction.

\paragraphb{Dataset Pre-Processing.}
In order to allow for a deep learning approach, we pre-processed our robot
movement dataset (Section~\ref{sec:dataset}) to remove fields with constant
values that would have no impact on the outcome (i.e. TCP flags which were the
same for all samples) and handle any NaN/null values.
We then normalised our dataset using the \textit{sklearn}~\cite{scikit-learn}
MinMaxScaler which scales and translates each feature individually to a real
number in the range of $(0, 1)$ on the training set. Due to the Y and YZ
movements having a larger distance range compared to other movements, there is a
larger sample size. Therefore, we also use \textit{sklearn} to stratify and
weight our dataset. We then used the {\em train\_test\_split} function to
produce our training and testing datasets. Specifically, during the randomised
splitting process, $20\%$ of samples were used for testing and from the training
set a further $20\%$ of the samples were used for validation. This resulted in a
60/20/20 split for training, testing and validation respectively.

\begin{figure}
  \centering
  \includegraphics[width=0.5\linewidth]{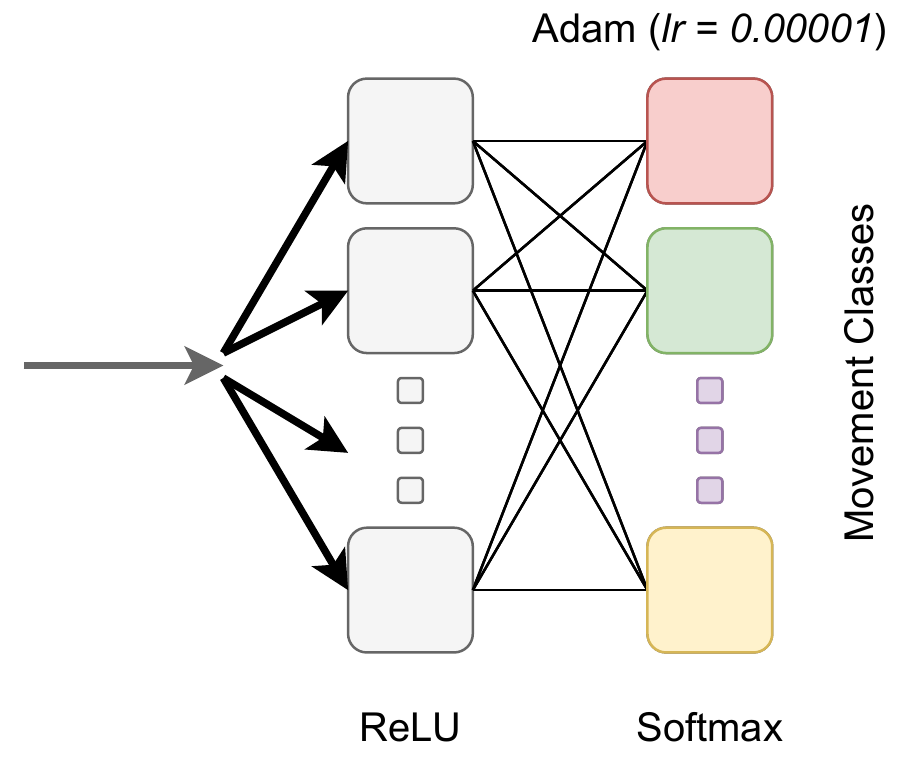}
  \captionsetup{singlelinecheck=off}
  \caption{
    \centering
    Neural Network Architecture
%     \hspace{\textwidth}{\small\em\textmd{This figure visually depicts the
% 	architecture and layers of our neural network used in the traffic analysis
% 	attack}}
  }
  \label{fig:model}
\end{figure}

\paragraphb{Deep Neural Network Architecture.}
After pre-processing the samples, we then constructed our neural network. To
create the network, we use the \textit{Keras}~\cite{chollet2015keras} Python
library to construct a sequential model (Figure~\ref{fig:model}). The input
layer consisted of 16 neurons for our 16 feature columns. After the input layer,
we use one hidden layer. This is a {\em Dense} layer with 108 neurons. The
choice of the number of neurons for this layer is based upon keeping the number
of neurons below $N$ to prevent over-fitting. We calculate $N$ using the
following formula~\cite{demuth2014neural}, for a single hidden layer:

\par
\begin{formula*}!{
	N_h = \dfrac{N_s}{(\alpha * (N_i + N_o))}
	}
    N_i: Number of input neurons,
	N_o: Number of output neurons,
	N_s: Number of samples in training dataset,
    \alpha: Arbitrary scaling factor,
\end{formula*}
\par

The alpha ($\alpha$) value is the effective branching factor (number of non-zero)
weights for each neuron, which we give a value of 2. The value of $N_s$ is 4968,
the number of samples in the training set for our baseline, such that we can
effectively compare against our other experiment parameters
(Section~\ref{sec:experimentparams}), giving us a maximum (optimum) number of
neurons 108.
% Since we should keep the number of neurons in this layer below $N$,
% we decided to use 32 neurons for our hidden layer, as previous tests (i.e. with
% 64 neurons) lead to overfitting.
This layer uses the ReLU activation function. Finally, the
output layer uses the SoftMax activation function~\cite{dunne1997pairing} to
have the output in the range of $[0,1]$ for use as predicted probabilities,
with categorical cross-entropy~\cite{zhang2018generalized} for our loss function.
We used Adam with a learning rate of $0.00001$ for our optimiser, with higher
learning rates resulting in lower accuracy scores. As we are using softmax
activation, this layer has 7 neurons with each corresponding to one of our
movement classes.

\paragraphb{\textit{Choice of Activation and Optimisation Functions.}}
We chose to use the ReLU activation function over other activation functions
as the reduced likelihood of vanishing gradient allows for a constant gradient
resulting in faster learning. Further, the sparsity of representations are shown
to be more beneficial than dense representations, as seen in other activations
such as sigmoids~\cite{krizhevsky2012imagenet,li2017convergence,agarap2018deep}.
We use the softmax activation function, combined with categorical
cross-entropy~\cite{zhang2018generalized} for our loss function as this is a
multi-class classification problem. Simply, a sample can belong to one of our
7 classes, with each class corresponding to one of the robot movements. As well
as this, we make use of the Adam optimiser -- an extension to the Stochastic
Gradient Descent (SGD) method based on adaptive estimation of first- and
second-order moments~\cite{kingma2014adam}. Specifically, it allows for the
updating of network weights iteratively based on the training data. We chose
this widely recommended optimiser as it fits best with our weighted sample sets
in opposition to other tried methods such as standard SGD, RMSProp and
SGD + Nesterov Momentum.

\section{Evaluation}

\subsection{Baseline Evaluation}

In accordance with our hypotheses, the first two ($RQ_1$ and $RQ_2$) aim to
determine whether we can detect the individual robot movements on each of our
robot's axes, as well as combinations of movement. Each movement corresponds to
the robot arm moving from some reference point to a destination in specified
directions. We term our first set of experiments the baseline, where the
distance of movement is set to 1mm and the lowest speed ($12.5mm/s$) with no
network parameters (link delay or packet loss) in effect. This allows us to
compare the impact of all parameters individually on classification accuracy.

\begin{table}[t]
\centering
\begin{tabular}{SSS}
	\toprule
	\textbf{Movement} & \textbf{Precision} & \textbf{Recall} \\
	\midrule
	X & 70\% & 85\% \\
	Y & 69\% & 54\% \\
	Z & 80\% & 63\% \\
	XY & 21\% & 60\% \\
	XZ & 68\% & 92\% \\
	YZ & 81\% & 31\% \\
	XYZ & 72\% & 97\% \\
	\bottomrule
\end{tabular}
\captionsetup{singlelinecheck=off}
\caption{
  \centering
  Baseline Classification Results
%   \hspace{\textwidth}{\small\em\textmd{The baseline samples
%   contain only samples with a single distance unit and lowest movement speed,
%   with no other varying parameters}}
}
\label{table:baselinematrix}
\end{table}

As depicted in Table~\ref{table:baselinematrix}, we can view the precision and
recall, as well as the classification accuracy for our
baseline~\cite{javaid2016deep}. The precision metric is the ratio of correctly
predicted positive movements to the total predicted positive movements. In the
case of our baseline, we can observe relatively good precision for most classes
averaging around \textasciitilde65\%, with the exception of the XY movement. % in the confusion matrix we can see that some movements are being incorrectly classed as XY or YZ movements.
Notably, the Z and YZ movements show the highest precision of around 80\%. In
the case of recall, the frequency of predicted movements are correctly
classified, which we can see is outstanding for the X, XZ and XYZ movements.
However, as with precision, the recall for the movements involving Y show to be
the lowest. From Figure~\ref{fig:tfeatures}, we can observe that this is likely
due to the similarities in traffic features across these movements. In the
context of this study, good or perfect recall is desirable but not the most
important if certain movements may be missed. We see that the recall across most
movements is good and thus, the combination of certain movements -- even with
some missing -- can still lead to the identification of entire workflows.

\subsection{Workflow Reconstruction}

From this, the next stage of our experiments focuses on our second hypotheses
($RQ_3$), to determine whether we can indeed reconstruct warehousing workflows
which are made up of combinations of these baseline movements. And, if so, to
what degree of accuracy can they be recovered? For this experiment, the focus
was on a subset of common warehousing workflows that involve robotic arms.
This includes: pick-and-place, push, pull and packing operations. We chose these
workflows, as they represent those that are unique and common operations to a
realistic warehouse which makes use of these
robots~\cite{tirian2013automation,kimura2015mobile,bogue2016growth,wang2020goods}.
Quantifying the accuracy of recovering these workflows allows us to meaningfully
demonstrate that an adversary could reveal daily operating environments within
logistics supply chains.

For these operations, we took inspiration for movement trajectories from
existing industrial robot datasets, such as the {\em Forward Dynamics Dataset
Using KUKA LWR and Baxter}~\cite{polydoros2016reservoir} for pick and place and
the {\em Inverse Dynamics Dataset Using KUKA}~\cite{rueckert2017learning}
for push/pull. At the heart of these workflows is the actual dynamic movements
themselves which may be aided by additional input (i.e. from sensors).
Ultimately, given that movement patterns are the primary factor which
establishes these workflows, it is reasonable to conduct our experiment on
reconstructing workflows from traffic patterns solely using position data.
In total, we have over 100 sets of test samples for each workflow with varying
speeds of movement, distances and directions to be evaluated using our neural
network approach.

\begin{figure}[h]
	\centering
	\includegraphics[width=0.75\linewidth]{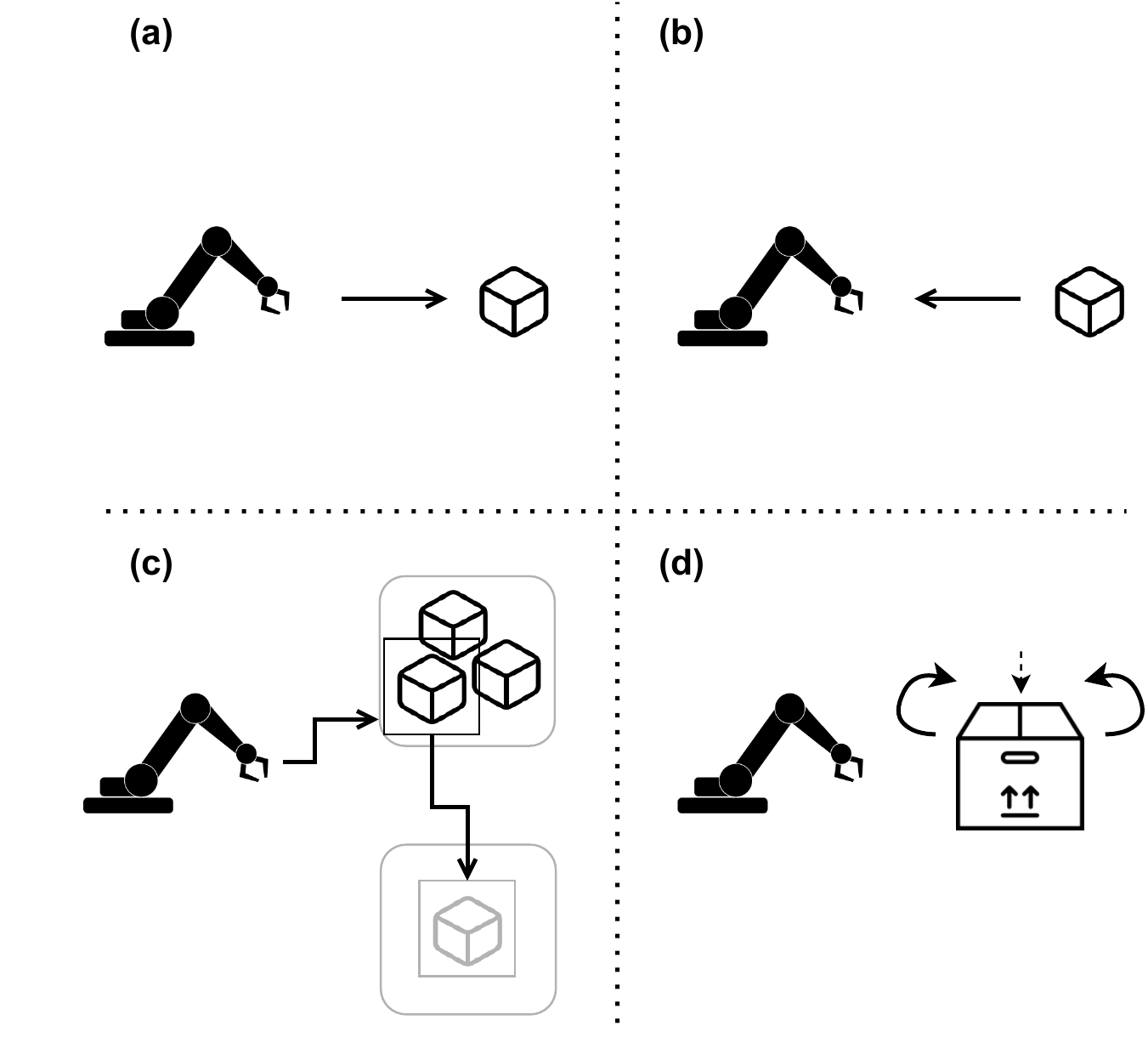}
	\caption{
      \centering
      Warehousing Workflows\hspace{\textwidth}{\small\em\textmd{
  % 	These results show the impact on classification accuracy with varying
  % 	distances for each movement. We find that
  	  (a) Push; (b) Pull; (c) Pick and Place; (d) Packing.}}
    }
	\label{fig:workflows}
\end{figure}

As shown in Table~\ref{table:recovery}, we can observe that on average the
manufacturing workflows can be recovered much better than individual movements,
averaging around 90\% accuracy. This is an important result as it demonstrates
that continuous monitoring of movement patterns can reveal potentially
confidential workflows and could be given to competing facilities, for example.
Further, this information can even combined with other side-channels such as
acoustic or power which may provide another level of granularity regarding
information leakage (i.e. identifying the weight of products being cross-docked
(pick-and-place) through acoustic reflections could potentially provide an
indicator for its contents).

\begin{table}[h]
\centering
\begin{tabular}{ccc} %c
	\toprule
	\textbf{Operation} & \textbf{Recovery Rate} & \textbf{Pos Changes} \\ % \textbf{Avg. Duration} & \textbf{Pos Changes} \\
	\midrule
	Push & 97\% & 2--3 \\ % 3.0s & 2--3 \\
	Pull & 97\% & 2--3 \\ % 3.0s & 2--3 \\
	Pick-and-Place & 84\% & 7--9 \\ % 7.4s & 7--9 \\
	Packing & 88\% & 6--9 \\ % 8.0s & 6--9 \\
	\bottomrule
\end{tabular}
\captionsetup{singlelinecheck=off}
\caption{
  \centering
  Workflow Reconstruction Results
%   \hspace{\textwidth}{\small\em\textmd{These results show
%   the rate (average accuracy across all sets per movement) at which the operation can be
%   successfully recovered.}}
}
\label{table:recovery}
\end{table}

\vspace{-2em}

\subsection{Impact of Experimental Parameters on Movement Classification}
\label{sec:experimentparams}

In our next set of experiments, we want to determine the impact of robot
parameters (distance and speed -- $RQ_4$) and network parameters
(jitter and packet loss -- $RQ_5$) on the classification accuracy. In real-world
networks the channel between the controller and robot may experience jitter,
packet losses or delays. Jitter and packet loss in real-world networks are known
to follow a Poisson
distribution~\cite{medhi2017network,muhizi2017analysis,dahmouni2009analytical,li2004performance,huremovic2014analytical},
particularly in the case of single-link communication (such as the case of
teleoperated robots) whose interaction can be abstracted to the likes of an
M/M/1 queue for example~\cite{mandjes1999end,kogel2013one}. To realistically
emulate link characteristics, we make use of link parameters in the Mininet SDN
for delay and packet loss between the robot and controller. Notably, Mininet
follows a Poisson process for packet arrival in a single link system, which
meets our expectations for real-world network emulation. Further, in our Mininet
SDN we use a link speed of 100Mb/s which has shown that emulating link
properties of delay and packet loss at this rate can be done
realistically~\cite{yan2017learning}. In the following experiments, we use a
base distance of $1mm$ and speed of $25k$ (corresponding to 12.5mm/s, with
future speed iterations being a factor of 2000 with a maximum speed of 100mm/s).

\begin{table*}[t]
  \centering
  \begin{tabular}{lcc|cc|cc|cc|cc|ccc}
    \toprule
	\multicolumn{14}{c}{{\scriptsize D = Distance (mm), P = Precision, R = Recall}} \\
	\midrule
    \multirow{2}{*}{} &
      \multicolumn{2}{c}{D=1} &
      \multicolumn{2}{c}{D=2} &
      \multicolumn{2}{c}{D=5} &
	  \multicolumn{2}{c}{D=10} &
	  \multicolumn{2}{c}{D=25} &
	  \multicolumn{2}{c}{D=50} & \\
      & {P} & {R} & {P} & {R} & {P} & {R} & {P} & {R} & {P} & {R} & {P} & {R} \\
      \midrule
    X & 70\% & 85\% & 60\% & 59\% & 62\% & 84\% & 38\% & 63\% & 36\% & 57\% & 52\% & 38\% \\
    Y & 69\% & 54\% & 76\% & 49\% & 69\% & 45\% & 75\% & 45\% & 54\% & 37\% & 67\% & 8\% \\
    Z & 80\% & 63\% & 72\% & 78\% & 74\% & 100\% & 30\% & 62\% & 31\% & 82\% & 41\% & 86\% \\
	XY & 21\% & 60\% & 28\% & 58\% & 38\% & 74\% & 68\% & 94\% & 67\% & 14\% & 26\% & 55\% \\
    XZ & 68\% & 92\% & 43\% & 74\% & 33\% & 35\% & 0\% & 0\% & 0\% & 0\% & 38\% & 65\% \\
    YZ & 81\% & 31\% & 67\% & 48\% & 80\% & 47\% & 72\% & 28\% & 96\% & 71\% & 84\% & 62\% \\
	XYZ & 72\% & 97\% & 78\% & 88\% & 61\% & 97\% & 36\% & 94\% & 54\% & 100\% & 53\% & 77\% \\
	% \midrule
	% \textbf{Accuracy} & \multicolumn{2}{c}{60\%} & \multicolumn{2}{c}{55\%} & \multicolumn{2}{c}{64\%} & \multicolumn{2}{c}{72\%} & \multicolumn{2}{c}{70\%} & \multicolumn{2}{c}{63\%} \\
    \bottomrule
  \end{tabular}
  \captionsetup{singlelinecheck=off}
  \caption{
    \centering
    Classification Results With Distance Parameter\hspace{\textwidth}{\small\em\textmd{
% 	These results show the impact on classification accuracy with varying
% 	distances for each movement. We find that
	Examining movement distance gives us a more fine-grained inferral of movements,\hspace{\textwidth}
	with classification accuracy increasing with movement distance in most cases
	compared to the baseline.}}
  }
  \label{table:distancematrix}
  \vspace{-2em}
\end{table*}

\subsubsection{Impact of Movement Distance on Classification Accuracy}

The first parameter we tested was the distance the robotic arm moved in a
particular direction. The results of this experiment can be seen in
Table~\ref{table:distancematrix}. For comparison, the baseline distance can be
observed at a single dstance unit ($D=1$). At 2 distance units (mm), we can
observe a decrease in precision among the X, Z, XZ and YZ movements. The X and Z
movements and XZ samples seem to be incorrectly classified as each other, with
most of the XZ samples being predicted actually being Y samples. Interestingly,
the low precision of XY is due to incorrect classifications of either Y or
YZ, perhaps due to the similarities in pack features between them. In most cases
we found an increase in recall. For the baseline, we already found
that movements involving the Y axis were lower in comparison with other
movements, but with this first increase in distance we find an increase in
their precision. At 5 distance units, we observe that the results are fairly
similar, with the XY movement precision increasing but with incorrect predictions
of Y and YZ movements as with 2 distance units. Notably, the Z movement has
perfect recall. With the XZ movement, we observe that the precision and recall
decrease further again, with most samples here being incorrectly as X or Z. At
10 distance units, we see decreases in precision and recall across all movements
aside from the Y and XY movements. At 25 units, we observe similar results to
those seen at 10 units, with the exception of YZ which has a precision of around
96\% and the recall of XYZ movements being perfect. Finally, at 50 distance
units, the XZ movement is classified correctly but not successfully, as seen at
lower distance units.

\paragraphb{Findings.} Overall we found that not only does the distance of
movement provide more granularity to inferred movements, but does also influence
the adversary's ability to fingerprint them. While an increase in distance does
reduce the accuracy of classifying some movements, we believe that changes in
the payload (i.e. larger number for distance in plaintext) and round trip time
better showcase trends on the Y-axis as distance increases. Further, among some
movement classes, we found that increasing the distance parameter does correlate
with incorrect predictions among similar classes (i.e. XZ with both the X and Z
movements).

\subsubsection{Impact of Movement Speed on Classification Accuracy}

For our next experiment, we look into how the speed of the movement affects the
classification accuracy, with the results depicted in
Table~\ref{table:speedmatrix}. For comparison, the baseline speed can be
observed at $S=25k$. At $S=50k$, we notice a slight decrease in precision for
most movement classes, aside from the Z and XY movements which show a slight
increase. Similar to the distance parameter, most XY predictions are actually
Y and YZ movement samples which we believe to be due to close similarities in
traffic features, however this is lower than the baseline leading to the slight
increase in precision. Similarly, we also observe a decrease in recall in the
same cases, aside from the X and XY movements. At $S=100k$, we observe further
decreases in precision and recall for the Y, XZ, YZ and XYZ movements. Notably,
the X movement has perfect precision and recall, and the Z and XY movements
show both improved precision and recall from the previous speed iterations.
At $S=150k$, we can see that the precision and recall for the X movement drop
to similar results as with $S=50k$. Similarly, among the Y, YZ and XYZ
movements we notice an increase in precision with others decreasing slightly
compared to previous results. The recall in most cases show a decrease, aside
from the YZ and XYZ movements which have the highest recall respectively
among all speed iterations. Finally, at $S=200k$, the results show to be similar
to the $S=100k$ results but with lower recall.

\paragraphb{Findings.} In this experiment we find that movement speed gives
improvements to movement classification compared to the baseline, specifically
at $S=100k$. We can also see that overall, we gain slightly better results for
fingerprinting movements are provided from the speed parameter when compared to
the distance parameter.

\begin{table*}[t]
  \centering
  \begin{tabular}{lcc|cc|cc|cc|ccc}
    \toprule
	\multicolumn{12}{c}{{\scriptsize S = Speed (mm/s), P = Precision, R = Recall}} \\
	\midrule
    \multirow{2}{*}{} &
      \multicolumn{2}{c}{S=25k} &
      \multicolumn{2}{c}{S=50k} &
      \multicolumn{2}{c}{S=100k} &
	  \multicolumn{2}{c}{S=150k} &
	  \multicolumn{2}{c}{S=200k} & \\
      & {P} & {R} & {P} & {R} & {P} & {R} & {P} & {R} & {P} & {R} \\
      \midrule
    X & 70\% & 85\% & 66\% & 89\% & 100\% & 100\% & 67\% & 88\% & 66\% & 85\% \\
    Y & 69\% & 54\% & 62\% & 46\% & 61\% & 28\% & 65\% & 34\% & 56\% & 45\% \\
    Z & 80\% & 63\% & 82\% & 60\% & 90\% & 98\% & 80\% & 62\% & 75\% & 65\% \\
	XY & 21\% & 60\% & 24\% & 70\% & 26\% & 76\% & 20\% & 46\% & 23\% & 41\% \\
    XZ & 68\% & 92\% & 61\% & 86\% & 37\% & 83\% & 34\% & 70\% & 42\% & 68\% \\
    YZ & 81\% & 31\% & 79\% & 28\% & 69\% & 24\% & 74\% & 45\% & 66\% & 39\% \\
	XYZ & 72\% & 97\% & 68\% & 95\% & 66\% & 97\% & 86\% & 97\% & 73\% & 97\% \\
	% \midrule
	% \textbf{Accuracy} & \multicolumn{2}{c}{60\%} & \multicolumn{2}{c}{70\%} & \multicolumn{2}{c}{73\%} & \multicolumn{2}{c}{61\%} & \multicolumn{2}{c}{69\%} \\
    \bottomrule
  \end{tabular}
  \captionsetup{singlelinecheck=off}
  \caption{
    \centering
    Classification Results With Speed Parameter\hspace{\textwidth}{\small\em\textmd{
    % These results show the impact on classification accuracy with varying
    % speeds of each movement (mm/s). We find that
    Examining the speed of movement gives us a more fine-grained movement fingerprint\hspace{\textwidth}
	with classification accuracy showing similar increases across some iterations as with distance.}}
  }
  \label{table:speedmatrix}
  \vspace{-1em}
\end{table*}

\subsubsection{Impact of Network Link Delay on Classification Accuracy}
\label{sec:delayresults}
The next step is to observe the impact of varying network parameters on the
accuracy of our fingerprinting attack. Specifically, we first look at the link delay
to observe the impact on classification over real-world network scenarios. In
Mininet, this parameter corresponds to the packet delay time over the
link -- in this case between the robot and controller. While it is possible to
emulate a series of random delays throughout movement transmissions (as delays
may typically be unpredictable in terms of magnitude in a realistic setting),
we wanted to observe the impact of packet delay times over a range of values that
may be considered reasonable for continued operation in safety-critical contexts.

\begin{table}[t]
  \centering
  \begin{tabular}{lcc|cc|cc|ccc}
    \toprule
	\multicolumn{10}{c}{{\scriptsize L = Link Delay, P = Precision, R = Recall}} \\
	\midrule
    \multirow{2}{*}{} &
      \multicolumn{2}{c}{L=10ms} &
      \multicolumn{2}{c}{L=50ms} &
      \multicolumn{2}{c}{L=100ms} &
	  \multicolumn{2}{c}{L=1s} & \\
      & {P} & {R} & {P} & {R} & {P} & {R} & {P} & {R} \\
      \midrule
    X & 64\% & 89\% & 99\% & 100\% & 100\% & 100\% & 98\% & 100\% \\
    Y & 100\% & 100\% & 100\% & 99\% & 100\% & 100\% & 100\% & 99\% \\
    Z & 80\% & 63\% & 100\% & 100\% & 100\% & 100\% & 100\% & 100\% \\
	XY & 99\% & 90\% & 100\% & 100\% & 100\% & 100\% & 100\% & 100\% \\
    XZ & 89\% & 83\% & 76\% & 100\% & 100\% & 100\% & 100\% & 100\% \\
    YZ & 100\% & 99\% & 100\% & 85\% & 100\% & 100\% & 100\% & 100\% \\
	XYZ & 100\% & 99\% & 89\% & 100\% & 100\% & 100\% & 100\% & 100\% \\
	% \midrule
	% \textbf{Accuracy} & \multicolumn{2}{c}{99\%} & \multicolumn{2}{c}{100\%} & \multicolumn{2}{c}{100\%} & \multicolumn{2}{c}{100\%} \\
    \bottomrule
  \end{tabular}
  \captionsetup{singlelinecheck=off}
  \caption{
    \centering
    Classification Results With Network Link Delay\hspace{\textwidth}{\small\em\textmd{
    We observe a
	significant improvement when a low link delay is introduced, with the
	precision, recall and accuracy reaching perfect as the delay increases.}}
  }
  \label{table:delaymatrix}
\end{table}

First, we look at the impact of network link delay on movement classification,
with the results shown in Table~\ref{table:delaymatrix}. In comparison with the
baseline results shown in Table~\ref{table:baselinematrix}, we observe a
significant increase in both precision and recall in all cases, with the
majority of movements having perfect precision and recall. As seen in the
table, the X movement initially has the poorest precision and Z with the poorest
recall. In this case, a proportion of the X movements are incorrectly classified
as Z and vice-versa -- a similar trend seen in previous experiments. However,
overall as the delay increases we see significant improvements.

\paragraphb{Findings.} In this set of experiments, we find that introducing a
low link delay significantly improves the precision and recall of all movements.
As the delay increases over the robot-controller link, our results show that an
adversary can infer movements with some degree of link delay almost perfectly
where there are \textit{acceptable} delays, and even better with larger delays.
We note that this significant increase -- in comparison with our other
experiments and the baseline -- may be due to differences in round-trip time and
packet interarrivals for each movements, increasing the variation among them
collectively, unlike distance and speed which seem to only affect the payload of
the collected traffic.

\subsubsection{Impact of Network Packet Loss on Classification Accuracy}
\label{sec:lossresults}

Next, we look into the effect of network packet loss on classification of
movements. Realistically, failures or inefficiencies of network components that
carry the data, such as a faulty router or weak wireless signal, can cause lost
or dropped packets and thus should be accounted for. In a TLS connection,
\textit{TCP flow control} detects packet losses and attempts to retransmit these
messages for reliable communication and results in decreased throughput, which
we believe may have an impact on time series data gathered from our attack. In a
realistic scenario, the question surrounding potential and acceptable packet
losses are important. In many applications, quality of service considerations
are given based on the type of data sent. For a safety-critical IoT system such
as industrial robots, even the loss of some amount of packets could
result in delays that could lead to or be unable to prevent serious harm. It has
been noted that losses between 5\% and 10\% of the total packet stream seriously
impacts the quality of service~\cite{mansfield2009computer}. For completeness,
we start with a 10\% loss and move up to 50\% loss. While this may be rare and
potentially unavoidable if this is the case, we feel that it is still useful to
determine the feasibility of our attack. Even given a larger number of packet
loss, movement data may still be present with (spuriously) retransmitted
packets if the network was deemed unfit for a robot to continue reliably
performing operations over. In Mininet, the packet loss is the rate of packet
loss (\% of random packets per second) over a given link. The results for this
experiment can be seen in Table~\ref{table:packetlossmatrix}. In comparison with
the baseline, we observe similar results with the effect of network link delay
shown in Table~\ref{table:delaymatrix} with a significant increase across all
movements. At 10\% loss, we observe near perfect accuracy of the model with
only slight drops in precision most notably for the X movement. This is due to
some X samples being mis-predicted as XZ movements. At 25\% packet loss, we
see much more of an improvement with most classes having 100\% precision and
recall. The X movement here improves, however the Z movement precision decreases
in comparison to that at 10\% loss, where some predicted XZ movements are Y
movements. Finally, at 50\% loss, we again see similar trends. However, the
precision for the X, Z and XZ movements decreases but still notably better than
the baseline results.

\paragraphb{Findings. } As with our other network parameter experiment (link
delay), introducing a percentage packet loss over the robot-controller link also
results in greater precision, recall and overall classification accuracy of our
robot's movements in all cases. Given the use of TLS as our secure channel
technology for the robot-controller link in the emulated network, drops in
packet arrivals will result in transmissions with increased interarrival times.
Notably, some work highlights a possible correlation between packet loss and
higher link utilisation which can increase packet interarrival times~\cite{varga2006analyzing}.

\begin{table}[h]
  \centering
  \begin{tabular}{lcc|cc|ccc}
    \toprule
	\multicolumn{8}{c}{{\scriptsize L = Packet Loss, P = Precision, R = Recall}} \\
	\midrule
    \multirow{2}{*}{} &
      \multicolumn{2}{c}{L=10\%} &
      \multicolumn{2}{c}{L=25\%} &
      \multicolumn{2}{c}{L=50\%} & \\
      & {P} & {R} & {P} & {R} & {P} & {R} \\
      \midrule
    X & 86\% & 100\% & 100\% & 100\% & 92\% & 100\% \\
    Y & 100\% & 100\% & 100\% & 88\% & 100\% & 100\% \\
    Z & 100\% & 100\% & 100\% & 100\% & 84\% & 100\% \\
	XY & 100\% & 100\% & 100\% & 100\% & 100\% & 90\% \\
    XZ & 91\% & 83\% & 73\% & 100\% & 89\% & 83\% \\
    YZ & 99\% & 97\% & 100\% & 100\% & 100\% & 97\% \\
	XYZ & 100\% & 99\% & 100\% & 100\% & 100\% & 100\% \\
	% \midrule
	% \textbf{Accuracy} & \multicolumn{2}{c}{100\%} & \multicolumn{2}{c}{100\%} & \multicolumn{2}{c}{100\%} \\
    \bottomrule
  \end{tabular}
  \captionsetup{singlelinecheck=off}
  \caption{
    \centering
    Classification Results With Network Packet Loss\hspace{\textwidth}{\small\em\textmd{
    % These results show the impact on classification accuracy with varying packet
	% loss over the robot-controller link.
	We observe near perfect accuracy in most cases, with the precision and recall increasing as the loss increases.}}
  }
  \label{table:packetlossmatrix}
\end{table}

\vspace{-2em}

\subsection{Open-World Evaluation}

In our main set of experiments, we conducted closed-set testing where we assume
the attacker to detect a known, strict set of movements. However, realistically,
the attacker may only know a subset. Thus, this naturally provokes the need to
understand the impact on classification when only some movements are known. In
this case, a comprehensive approach for our dataset is labelling each movement
progressively and leaving the rest of the movements unlabelled, to observe the
impact of increasing numbers of unlabelled classes.

\begin{table*}
  \centering
  \begin{tabular}{lcc|cc|cc|cc|ccc}
    \toprule
	\multicolumn{12}{c}{{\scriptsize U = \# Unknowns, P = Precision, R = Recall}} \\
	\midrule
    \multirow{2}{*}{} &
      \multicolumn{2}{c}{U=2} &
      \multicolumn{2}{c}{U=3} &
      \multicolumn{2}{c}{U=4} &
	  \multicolumn{2}{c}{U=5} &
	  \multicolumn{2}{c}{U=6} & \\
      & {P} & {R} & {P} & {R} & {P} & {R} & {P} & {R} & {P} & {R} \\
      \midrule
    X & 68\% & 84\% & 73\% & 86\% & 71\% & 87\% & 52\% & 99\% & 54\% & 99\% \\
	Y & 68\% & 55\% & 71\% & 65\% & 62\% & 79\% & 61\% & 82\% & & & \\
	Z & 79\% & 63\% & 79\% & 71\% & 79\% & 65\% & & & & & \\
	XY & 23\% & 67\% & 25\% & 74\% & & & & & & & \\
	XZ & 67\% & 90\% & & & & & & & & & \\
	Unknown & 95\% & 54\% & 97\% & 62\% & 88\% & 74\% & 90\% & 65\% & 99\% & 92\% \\
	% \midrule
	% \textbf{Accuracy} & \multicolumn{2}{c}{92\%} & \multicolumn{2}{c}{75\%} & \multicolumn{2}{c}{80\%} & \multicolumn{2}{c}{60\%} & \multicolumn{2}{c}{63\%} \\
    \bottomrule
  \end{tabular}
  \captionsetup{singlelinecheck=off}
  \caption{
    \centering
    Open-World Classification Results\hspace{\textwidth}{\small\em\textmd{
    These results show the impact on classification accuracy when decreasing the
	number of known movements while increasing the number of unknown movements.
    We observe a correlation of decreased accuracy decreases as the number of
	known movements increase.}}
  }
  \label{table:unknownsmatrix}
  \vspace{-2em}
\end{table*}

As shown in Table~\ref{table:unknownsmatrix}, we present an open-set approach
which involves labelling each class progressively, leaving the rest unlabelled.
The aim of this is to observe the impact of n-unknown classes. Within this set
of experiments, we do not count 1 unknown class, as this would be the same
as our baseline. First, for 2 unknown classes ($U=2$), we see fairly
similar outcomes when compared to the baseline results shown in
Table~\ref{table:baselinematrix}. For 3 unknowns, we see improvements in
precision and recall for all remaining movements. At 4 unknowns, we only see
slight improvement for the Z movement, and the recall of the X and Y movements,
but a reduction in precision for X and Y compared to 3 unknowns. For 5 unknowns,
precision and recall for the Y movement stay relatively consistent, but the
precision for X drops greatly with the recall increasing. A greater percentage
of X movements are incorrectly predicted where they should fall into the unknowns.
Finally at 6 unknowns, we see similar results to 5 unknown classes but the
recall is near perfect for the X movement. Overall, we observe that the increase
in precision and recall for unknown movements increases and this is expected.
Furthermore, the increase in recall in most cases for movements still known is
also expected given that the subtle feature differences are more present given
a larger variance in feature values for the unknown set.

% Discussion
\section{Discussion}
\label{sec:discussion}

We have presented the results of experiments investigating whether an adversary
can identify robot movements, even when the traffic has been encrypted. In
our case, we investigated whether a movement can be identified from
preliminary analysis of traffic features such as flow length, and then provide a
shallow-learning approach to use these traffic features and select the best features
for classifying movements. We then further looked into the ability to reconstruct
procedures, specifically in the context of industrial warehouses.

From our preliminary analysis, we found that simply observing variations among
traffic features in our sample set one can identify
individual movements with some ease. However, similarities arise among
all traffic features for other movements (i.e. combined movements such as XY) which
makes them harder to identify in this manner (Figure~\ref{fig:tfeatures}). This
motivates the need for a deep learning approach to identify robot operations over
a TLS-secured channel.
% Fingerprinting attacks based on deep-learning techniques have
% been successful in other areas such as websites~\cite{oh2019p1} and VPN traffic~\cite{lotfollahi2020deep}.
% Thus, we investigate how effective a similar attack would be for robotic systems.

Our evaluation shows that it is indeed possible
for a passive attacker to reconstruct procedures even when the traffic is
encrypted under TLS, with high accuracy. As a baseline, we show that most movements
which our robot can carry out can be predicted with at least 60\% accuracy and
precision. However, in a realistic setting this may be less than ideal. For example,
let's consider MIT and Boston Dynamic's \textit{Dr. Spot}~\cite{drspot,huang2020agile},
which makes use of a teleoperated quadruped robot for measuring patient vital signs.
Our work demonstrates that parameters such as distance, for example in the
case of the quadruped robot from patient
and angle of view, could reveal what vitals are being measured. In combination with
other sources of information such as GPS information, this may point to what triage
zones were visited. This provoked us to explore another level of
granularity of movement, specifically in our robot -- the distance and speed of
movements. In our results, we show that these parameters are meaningful to an attacker,
with the accuracy and precision increased by at least 10\%. Furthermore, in realistic
case settings such as Dr. Spot, teleoperation would be conducted over a wireless network
where factors such as packet loss and delay/latency come into play. In our experiments,
we show that under such realistic teleoperating conditions, an attacker can identify robot movement
operations with very high accuracy.

We take this a step further and demonstrate that we can identify a higher-level
of abstraction. Specifically, identifying warehousing workflows (i.e. pick-and-place)
where the intuition behind this shows promise to ultimately piecing
together entire manufacturing workflows. From here, it could then be possible to
compromise operational confidentiality by combining information from this
attack with other sources (such as delivery times, package meta-data, information
from other side-channels, etc.).
% Furthermore, our robot only provides a minimal
% set of movement operations. Realistically, other industrial and surgical robots
% may employ a number of actuators (whereas our robot makes use of one) and a number
% of sensing devices, which would provide not only more information for use in
% classifying movements, but also allow for more operations to be conducted compared
% to our robot which makes use of 7 fundamental movement operations.

\paragraphb{Emulation of Network Link Characteristics}
For measuring the impacts of link characteristics such as packet loss
and delay/jitter, a suitable network emulation environment is required. We
consider several options such as PlanetLab~\cite{chun2003planetlab}, DETER~\cite{benzel2007design,benzel2009current}, Emulab~\cite{johnson2006mobile},
NS3~\cite{riley2010ns} and Mininet~\cite{kaur2014mininet,fontes2015mininet}.
We chose Mininet as it is best suited for
modelling arrivals as a Poisson process, a similar behaviour as in teleoperated robots~\cite{paxson1994wide}.
The justification for not using other approaches is as follows.
PlanetLab~\cite{chun2003planetlab} is a global testbed for network systems
research with nodes spread across the earth. The main challenge associated with
this is that results are not reproducible as network conditions can vary over
time~\cite{spring2006using}. Other real-world emulation testbeds such as
DETER~\cite{benzel2007design,benzel2009current} and Emulab~\cite{johnson2006mobile} face similar challenges, where
resources are shared among many researchers which can skew results.
NS3~\cite{riley2010ns} on the other-hand is designed as a discrete-event network simulator
and thus, is therefore unsuitable for this work as the network layer will be
simulated even if the robot-edge runs live. While Mininet is not without limitations -- primarily
scalability (due to single threaded components in the core) -- in our case
of a small network with a few nodes and switches, these limitations do not
manifest. Ultimately, it would be interesting to observe how other arrival
models for delay and loss would impact movement classification~\cite{chydzinski2020queues,moscholios2019efficient}.

\begin{table}[h]
\centering
\begin{tabular}{SSS}
	\toprule
	\textbf{Movement} & \textbf{Precision} & \textbf{Recall} \\
	\midrule
	X & 38\% & 57\% \\
	Y & 53\% & 88\% \\
	Z & 76\% & 8\% \\
	XY & 35\% & 45\% \\
	XZ & 86\% & 49\% \\
	YZ & 44\% & 27\% \\
	XYZ & 60\% & 80\% \\
	\bottomrule
\end{tabular}
\captionsetup{singlelinecheck=off}
\caption{
  \centering
  Tor Classification Results
%   \hspace{\textwidth}{\small\em\textmd{The Tor samples make use
%   of samples from the same characteristics as our baseline study, showing a
%   classification accuracy of \textasciitilde49\%.}}
}
\label{table:tor}
\end{table}

\vspace{-1em}

\subsection{Limitations}

% Applying our traffic analysis approach leads to some natural limitations which are a function
% of the patient, robot, and toolkit. We find two key limitations, however further study is required
% which may highlight others and compensation techniques for these.

% \paragraphb{Disconnect Between Tool Motion and Procedure.} Our approach achieved significant accuracy
% of reconstruction, but it contains a number of limitations. One limitation arises from the deviation
% between robot path (path of arm/tool) and the procedure carried out. When the surgical tools are
% deployed they may have the same traffic pattern for many operations. For instance, during a scalpel
% operation, its thickness may result in different cuts for the same position/direction. Further, the
% exact starting position of the surgery can have significant impact on planning, as tools must account
% for characteristics (i.e. physical height) of the area of operation.
% As another example the cutting rate in case of incisions or suturing rate in case of stitching is
% also a key parameter which is unlikely to be recovered from a traffic analysis approach. Appropriate
% compensation techniques can mitigate these effects but largely these parameters are not derivable
% from traffic analysis.

\paragraphb{Ageing Effects.} Our attack is evaluated on a single robot. However, one
question which arises is that if the classifier is trained on a new unit, would
it be possible to demonstrate whether the classifier deteriorates over time as the
robot ages? In the case of our fingerprinting attack via TLS traffic analysis,
this should be mostly unaffected with the payload information staying the same
throughout its age. However, network component ageing may have an impact over
time on possible link delays and losses, for example, with this demonstrated in
Section~\ref{sec:delayresults} and \ref{sec:lossresults}. In any case, to address
the concept of ageing in the case of the robot itself, the exploration of
other side channels (such as acoustics and power analysis) could be studied to
observe the impact of ageing on movement fingerprinting. This is due to the
fact that side-channels based on physical processes are subject to ageing, such
as with belt-driven robots developing slack over time which can induce drift in
physical characteristics (i.e. noise). Further, the impact of device calibration
over time could result in variations of robot movements (such as uncalibrated motor
axes which can result in unpredictable degrees of movement).

\subsection{Countermeasures (Tor)}

Tor is a potential countermeasure to fingerprinting robot
operations via traffic analysis. Tor is a low-latency, circuit-based overlay
network which enables anonymous communication by allowing different streams to
overlap each other such that traffic volumes will still be hidden. Given that
Tor shows some success as a defence, such as in website fingerprinting~\cite{juarez2015wtf},
it is important to see how Tor performs as a countermeasure in the context of
our robot. To do this, we setup a Tor hidden service which receives the control
commands sent by our controller over HTTPS to the hidden service. Using Wireshark,
we were able to monitor incoming traffic to the hidden service machine and
capture common flows which correspond to the robot traffic. Within this
experiment, the traffic was routed through multiple autonomous systems over
around 20 circuits. From the results in Table~\ref{table:tor},
we can see that in comparison with our baseline results shown in
Table~\ref{table:baselinematrix}, the precision across most movements decreases
slightly (averaging \textasciitilde20\% decrease), with the exception of the XY
and XZ movements which show an increase in precision of 14\% and 18\% respectively.
In the case of recall, we also notice more decreases across the movements, with
an increase of 34\% only present for the Y movement. Notably, we also notice a
substantial drop in recall for the Z and XZ movements with a decrease of 55\%
and 43\% respectively. Further, looking into performance impact of Tor, we can
see that the latency does not (overall) present as a big problem for many cases
with packet times fairly sporadic but under 1s. However, in critical cases
cases this wait time for control commands to be received may be less than
desirable. We then explored the success of workflow reconstruction when Tor
is employed as a countermeasure. The results are described in Table~\ref{table:torrecovery}.
In comparison with our reconstruction results found in Section~\ref{sec:recon},
we find that the use of Tor as a countermeasure reduces the recovery rate by at
least a factor of 2, which is a significant drop. However, this accuracy in
comparison with baseline invidual movements (non-patterns) is relatively similar.
% Thus, as a whole, while Tor may seem suitable as a defence to some extent for
% inhibiting workflow reconstruction, other countermeasures may provide a more
% suitable defence against our attack as a whole.

\begin{table}[h]
\centering
\begin{tabular}{ccc} %c
	\toprule
	\textbf{Operation} & \textbf{Recovery Rate} & \textbf{Pos Changes} \\ % \textbf{Avg. Duration} & \textbf{Pos Changes} \\
	\midrule
	Push & 45\% & 2--3 \\ % 3.0s & 2--3 \\
	Pull & 47\% & 2--3 \\ % 3.0s & 2--3 \\
	Pick-and-Place & 51\% & 7--9 \\ % 7.4s & 7--9 \\
	Packing & 48\% & 6--9 \\ % 8.0s & 6--9 \\
	\bottomrule
\end{tabular}
\captionsetup{singlelinecheck=off}
\caption{
  \centering
  Tor Procedure Reconstruction Results
%   \hspace{\textwidth}{\small\em\textmd{The results show that
%   procedure reconstruction is hindered when Tor is employed, in comparison
%   with those shown in Table~\ref{table:recovery}.}}
}
\label{table:torrecovery}
\end{table}

\vspace{-2em}

\begin{figure}
\centering
\begin{subfigure}{\linewidth}
  \centering
  \includegraphics[width=\linewidth]{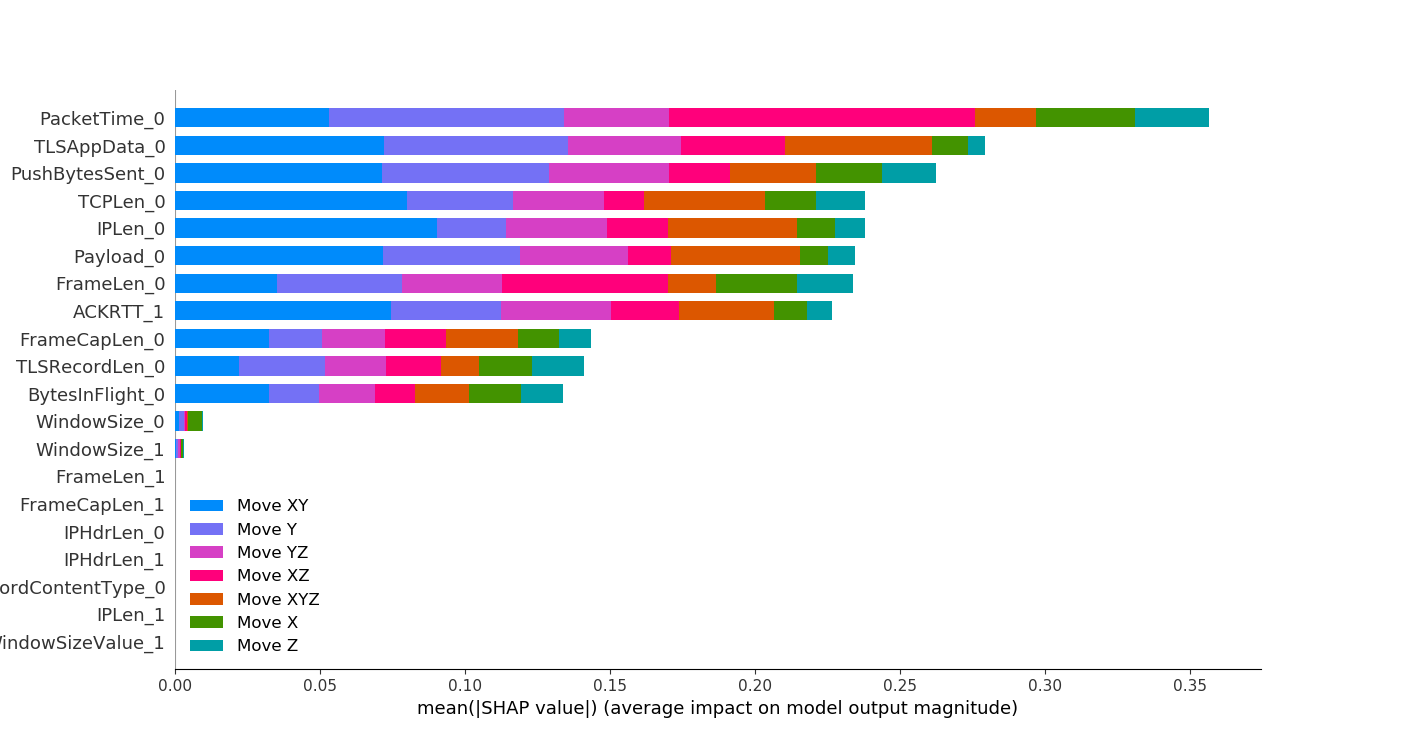}
  \caption{SHAP -- TLS}
  \label{fig:shaptls}
\end{subfigure}
\begin{subfigure}{\linewidth}
  \centering
  \includegraphics[width=\linewidth]{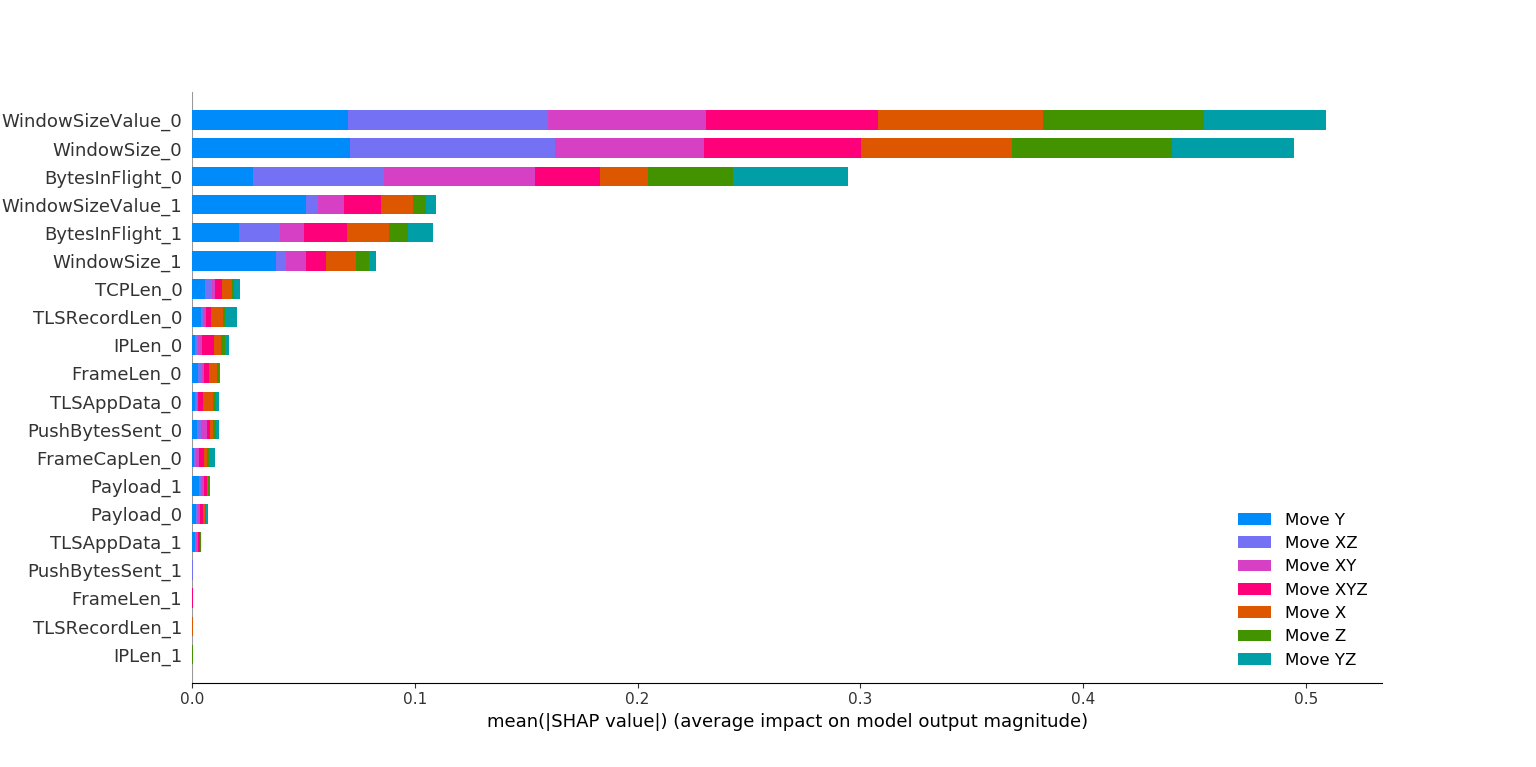}
  \caption{SHAP -- Tor}
  \label{fig:shaptor}
\end{subfigure}
\captionsetup{singlelinecheck=off}
\caption{\centering SHAP Values for TLS/Tor Datasets\hspace{\textwidth}{\small\em\textmd{
	% ...
}}}
\label{fig:shap}
\end{figure}

Given that the use of Tor as a countermeasure shows to be successful, it is
interesting to dive deeper and understand why exactly this is the case. In
Figure~\ref{fig:shap}, we observe the SHAP values~\cite{lundberg2017unified} for analysing the
most prominent features in both the TLS and Tor datasets. SHAP values allow for
us to evaluate the impact of our features on the predictions made by our model.
We can see that without the Tor countermeasure employed (Figure~\ref{fig:shaptls}),
the packet time remains the most prominent feature, with TCP packet features
such as application data, header length and payload considered highly important
to the neural network. However, under Tor, it is interesting to observe that
neither of these features are of much importance to the network for classification.
Instead, features such as the window size and bytes in-flight are the key
features. % From these observations, we question why this is the case.

In the case of TLS packet-size related features, Tor connections
make use of padding cells sent in both directions at varying
transmission intervals depending on consensus parameters. This
leads to the payload of TCP packets to be tranmitted in fixed-size
cells of 514 bytes, or if the payload is smaller then the cell is
zero-padded. While there may be some similarities between movement
patterns, with regard to existing TCP features, a lack of variability leads to
lowered reliance by the neural network. Interestingly, Tor does not affect the
bytes in flight or the window size. The BytesInFlight feature is defined by
Wireshark as an indicator for the amount of unacknowledged data that our
controller has transmitted. % It is always less than or equal to the recipient's receive window.
% How is bytes-in-flight not affected
The shorter the {\em distance} for receiving ACKs (faster time) results in lower
bytes in-flight, and ultimately a lowered window size needed for optimal
performance.
The window size is an advertisement from the receiving robot of how many bytes
of data it can receive at some point in time to control data flow, which may be
dependent on the movement(s) being carried out. In Tor, the data is in equal-sized
cells which leave the window size to be constant as multiple tor circuits are
multiplexed through the same TCP
connection~\cite{calvo2020classifying,tschorsch2016mind,fischer2011privacy}.
In our experiments we use different circuits for each movement and conducted
multiple runs for each movement on different circuits to minimise this, however
window size seems to remain an important feature.

Overall, we find Tor does reduce the precision and recall in most cases which
lowers success of fingerprinting movements. However, given that some decreases
are only slight and a presence of increase in cases, it may still be possible
for an attacker to fingerprint movements. Under Tor however, we find that the
procedure reconstruction rate is greatly reduced, leaving little success left
for an attacker to reconstruct operations.
% While only slight decreases are present
% when Tor is employed looking at movements individually, it is interesting to
% see that, in comparison, the reconstruction of surgical operations is a much
% more challenging task.
Future work related to this could be to conduct a similar study on different
robots and communication architectures (such as the P2P subscriber model
employed by ROS~\cite{quigley2009ros}) to better generalise the impact of
this attack across robotics systems.

\subsection{Future Work}

As a part of future work, we consider two other potential countermeasures: (1)
padding robot traffic and (2) mixing robot traffic with other background traffic.

\paragraphb{Padding.} First, we consider padding the robot traffic
in an attempt to counter our proposed traffic analysis attack, with the idea based
on the perfect secrecy theory proposed by Shannon~\cite{shannon1949communication}
(constant-rate padding). Constant-traffic-rate techniques (inserting dummy packets
to create padded traffic) have been shown to not prove as effective against
statistical traffic analysis techniques, and in particular for our case which
makes use of the traffic time-series features to detect movements. Fu et
al.~\cite{fu2003countermeasures} propose a variant rate traffic padding
countermeasure which can defend against attacks by leveraging sample variance and
entropy to exploit correlations between traffic rate and packet
inter-arrival times of padded traffic. Other approaches such as BuFLO~\cite{dyer2012peek} which removes side-channel
information by sending packets of fixed-length at fixed intervals also show
promise for areas such as HTTP traffic analysis. However in the case of many
safety-critical robotics systems which are time-critical, this may be less than
ideal. In any case, it would be interesting to determine the impact of a countermeasure that
pads traffic using variable inter-arrival times, which
have shown to be effective regardless of sample statistics collected by an
adversary when sample distributions of the inter-arrival times for our robot's
traffic are analysed to produce a design guideline for a VIT-based
approach~\cite{fu2003countermeasures,bushart2020padding}.

\paragraphb{Mixing.} Second, one can consider mixing in background
traffic with the robot traffic. With some work demonstrating that approaches
for regularizing traffic (i.e. constant-rate padding) and confusing the adversary,
many require a high data overhead (more packets) or induce delays in general
traffic, which question the suitability as a defense in a time-critical context
such as industrial robotics. Instead, a more lightweight approach which does not
require any additional infrastructure would be ideal. In this case, we look at
the potential for mixing in background traffic with the robot traffic, such
as GLUE~\cite{gong2020zero} which adds in dummy traces (each corresponding to a
single webpage) to have DNS traffic appear to the attacker as a longer consecutive
trace. This makes it harder to identify end points with minimal overhead.
This is shown to be more successful than existing defenses in the area
of website fingerprinting, as many existing attacks rely on single traces.
Ultimately, it would be interesting to determine if adding dummy movement traces
has an impact on movement classification.

\paragraphb{Other Fingerprinting Attack Vectors.}
As well as countermeasures, it would be interesting to consider other vectors
which could contribute to the classification of robot movements, such as the
ability to fingerprint sensor information. If an attacker was able to fingerprint
sensor information would help to provide more fine grained information to aid in
detecting operations. For example, a series of movement operations could relate
to a pick-and-place operation, but the weight of packages and other metadata
could be inferred from sensor outputs, which in combination with
classified movement operations could paint a more detailed picture. Further,
the Robot Operating System
(ROS)~\cite{quigley2009ros,mcclean2013preliminary,demarinis2019scanning} is
becoming more prominent in robotics research and is envisioned to be an upcoming
standard for robot middleware. Compared to the robot architecture we use in this
work, ROS-based systems use a publisher-subscriber model where sensors and other
input devices will act as publishing nodes and provide data to actuators
(subscriber nodes), where the communication is peer-to-peer (P2P). Another point
for future work would be to analyse the traffic flowing within and out of
ROS-based systems, and compare results with this study.

% \subsection{Open-World Recognition}
%
% In open-world recognition (OWR), a recognition system should be able to: (1)
% detect unseen information (also having no semantic relation to training data),
% (2) select samples that will be added to the existing model, (3) label these
% samples and (4) update the classifier~\cite{geng2020recent}. In comparison,
% this is an extension to open-set recognition where known classes are trained on
% and unseen classes are simply identified and nothing else. While it may be useful
% to adopt an open-set approach on top of this work, it will provide little benefit
% to the adversary as a manual approach on top of this would be required to understand
% the unknown data. Thus, an open-world approach would be much more beneficial.
% However, the authors in \cite{geng2020recent} describe that while recent methods
% have been proposed for the open world framework such as Open Deep Networks~\cite{shu2018odn},
% incremental learning~\cite{rudd2017extreme} and meta-learning~\cite{xu2019open},
% there is much more work to be carried out before it can be applied practically.

% Related Work
\section{Related Work}
\label{sec:related}

In this work, we determine whether we can detect robot movement operations
from traffic patterns, when the channel is protected by TLS. The problem of
detection and prediction of applications from encrypted traffic traces has been
investigated from a variety of different angles. Early approaches to identification
and prediction focused on identifying application traffic such as for firewalls,
websites and quality of service mechanisms and identifying actions (such as device user
actions)~\cite{alshammari2011can,velan2015survey}. Many of these early approaches
focus on payload-based classification and signature-based detection, which do
not work as well when traffic is encrypted~\cite{sen2004accurate}. Further,
such approaches took either a graphical approach -- through understanding of
social networks and motifs to understand patterns of communication and relationships
between features -- or a simple statistical approach through probability density
functions of traffic features and port-based classification~\cite{khalife2014multilevel,velan2015survey}.
From these limitations, the advent of machine learning approaches have shown to be
advantageous in relieving such limitations, combining statistical and graphical
approaches to build patterns to associate traffic with application
protocols~\cite{moore2005internet}. However, machine learning approaches mainly
apply to labeled data and require to be ``taught'' when results are incorrect,
to which deep neural networks do not require human intervention to learn from
mistakes (machine learning approaches almost always require structured data).

With the ubiquitous nature of network traffic, the prominence of deep learning
techniques increased significantly in areas such as: traffic classification deep learning
methods used for website~\cite{oh2019p1}; device/user~\cite{xu2015device,abe2016fingerprinting}
fingerprinting; and distinguishing between VPN and non-VPN traffic~\cite{lotfollahi2020deep},
among others~\cite{aceto2018mobile,rezaei2019deep,yang2018tls}. While many of
these techniques have shown success, there has been little on identification of
safety-critical IoT systems such as robotics which is the focus of this work.
Oh et al.~\cite{oh2019p1} describe the use of deep neural networks for predicting
website fingerprinting and show that in comparison to other state-of-the-art
fingerprinting techniques, the deep learning approach demonstrated a significant
increase in classification accuracy. Furthermore, the promise of deep learning
approaches is backed up by similar successes in identification over TLS-encrypted
traffic~\cite{aceto2018mobile,rezaei2019deep,yang2018tls}. Further, these similar
approaches describe that the additional features supplied to traffic with the
adoption of TLS can bring higher accuracy compared to standard packet features
used in earlier approaches, such as packet size and timing features.

% Conclusion
\section{Conclusion}
\label{sec:conclusion}

In conclusion, we present a case for evaluating whether a passive adversary can
still identify robot movements, even when the traffic between a robot and
controller (in a teleoperated architecture) is encrypted under TLS. We propose
a deep learning approach, which shows that it is possible for an adversary to
successfully classify our robot's movements when protected by TLS with around
60\% accuracy. Furthermore, we demonstrate that taking into account more fine-grained
movement details such as distance of movement the accuracy increases, and
when factors that impact network traffic, such as packet loss and link delay,
are taken into account, we can achieve perfect accuracy (100\%) for classifying
our robot's movements.

\begin{acks}
%\section{Acknowledgements}
	The authors are grateful for the support by the Engineering and Physical Sciences
	Research Council (11288S170484-102), the PETRAS Centre for IoT Cybersecurity, and the support of the
	National Measurement System of the UK Department of Business, Energy \&
	Industrial Strategy, which funded this work as part of NPL's Data Science program.
	% , UKIERI-2018-19-005, and...
\end{acks}

% References
\bibliographystyle{ACM-Reference-Format}
\bibliography{references}

\end{document}